\documentclass[12pt]{article}  
\usepackage{epsf}
\usepackage{graphics}
\usepackage{psfig}
\catcode`\@=11

\@addtoreset{equation}{section}
\def\theequation{\arabic{section}.\arabic{equation}}
    
\catcode`\@=11

\def\section{\@startsection{section}{1}{\z@}{3.5ex plus 1ex minus
   .2ex}{2.3ex plus .2ex}{\large\bf}}

\def\eqnarray{\let\@currentlabel=\theequation\refstepcounter{equation}
    \global\@eqnswtrue
    \global\@eqcnt\z@\tabskip\@centering\let\\=\@eqncr
    $$\halign to \displaywidth\bgroup\@eqnsel\hskip\@centering
      $\displaystyle\tabskip\z@{##}$&\global\@eqcnt\@ne
       \hfil${{}##{}}$\hfil
      &\global\@eqcnt\tw@ $\displaystyle\tabskip\z@{##}$\hfil
       \tabskip\@centering&\llap{##}\tabskip\z@\cr}
\def\lefteqn#1{\hbox to 4\arraycolsep{$\displaystyle #1$\hss}}
\def\thesection{\arabic{section}.}

\def\appendix{\setcounter{section}{0}
        \def\thesection{Appendix.}
        \def\theequation{\Alph{section}.\arabic{equation}}}

\long\def\@makefntext#1{\parindent 0cm\noindent
\hbox to 1em{\hss$^{\@thefnmark}$}#1}
\def\IR{{\hbox{{\rm I}\kern-.2em\hbox{\rm R}}}}
\def\IH{{\hbox{{\rm I}\kern-.2em\hbox{\rm H}}}}
\def\IC{{\ \hbox{{\rm I}\kern-.6em\hbox{\bf C}}}}
\def\IZ{{\hbox{{\rm Z}\kern-.4em\hbox{\rm Z}}}}

\newcommand{\beq}{\begin{equation}}
\newcommand{\be}{\begin{equation}}
\newcommand{\eeq}{\end{equation}}
\newcommand{\ee}{\end{equation}}
\newcommand{\bea}{\begin{eqnarray}}
\newcommand{\eea}{\end{eqnarray}}
\newcommand{\bean}{\begin{eqnarray*}}
\newcommand{\eean}{\end{eqnarray*}}
\newcommand{\ba}{\beq\begin{array}{lll} }
\newcommand{\ea}{\end{array}\eeq}

\def\IC{ {\rm l\hspace{-1.2ex}C} }    
\def\IZ{{\hbox{{\rm Z}\kern-.4em\hbox{\rm Z}}}} 
\def\IR{{\hbox{{\rm I}\kern-.2em\hbox{\rm R}}}} 

\begin{document}                           %
                                                                 %
                                                                 %
%
%
%
%
\def\citen#1{%
\edef\@tempa{\@ignspaftercomma,#1, \@end, }
\edef\@tempa{\expandafter\@ignendcommas\@tempa\@end}%
\if@filesw \immediate \write \@auxout {\string \citation {\@tempa}}\fi
\@tempcntb\m@ne \let\@h@ld\relax \let\@citea\@empty
\@for \@citeb:=\@tempa\do {\@cmpresscites}%
\@h@ld}
%
\def\@ignspaftercomma#1, {\ifx\@end#1\@empty\else
   #1,\expandafter\@ignspaftercomma\fi}
\def\@ignendcommas,#1,\@end{#1}
%
%
\def\@cmpresscites{%
 \expandafter\let \expandafter\@B@citeB \csname b@\@citeb \endcsname
 \ifx\@B@citeB\relax 
    \@h@ld\@citea\@tempcntb\m@ne{\bf ?}%
    \@warning {Citation `\@citeb ' on page \thepage \space undefined}%
 \else
    \@tempcnta\@tempcntb \advance\@tempcnta\@ne
    \setbox\z@\hbox\bgroup 
    \ifnum\z@<0\@B@citeB \relax
       \egroup \@tempcntb\@B@citeB \relax
       \else \egroup \@tempcntb\m@ne \fi
    \ifnum\@tempcnta=\@tempcntb 
       \ifx\@h@ld\relax 
          \edef \@h@ld{\@citea\@B@citeB}%
       \else 
          \edef\@h@ld{\hbox{--}\penalty\@highpenalty \@B@citeB}%
       \fi
    \else   
       \@h@ld \@citea \@B@citeB \let\@h@ld\relax
 \fi\fi%
 \let\@citea\@citepunct
}
%
\def\@citepunct{,\penalty\@highpenalty\hskip.13em plus.1em minus.1em}%
%
%
\def\@citex[#1]#2{\@cite{\citen{#2}}{#1}}%
%
%
\def\@cite#1#2{\leavevmode\unskip
  \ifnum\lastpenalty=\z@ \penalty\@highpenalty \fi 
  \ [{\multiply\@highpenalty 3 #1
      \if@tempswa,\penalty\@highpenalty\ #2\fi 
    }]\spacefactor\@m}
\let\nocitecount\relax  
\title{4D Tropospheric Tomography using GPS Estimated  Slant Delays}
\author{A. Flores\thanks{Institut d'Estudis Espacials de Catalunya (IEEC), CSIC Research Unit
Edif. Nexus-204, {Gran Capit\`{a}} 2-4, 08034 Barcelona, Spain.
e-mail: flores@ieec.fcr.es}, G. Ruffini and A. Rius}
\maketitle

\abstract
Tomographic techniques are successfully
applied to obtain 4D images of the tropospheric refractivity
in a local dense network. 
In the lower atmosphere both the small height and time scales 
and the non-dispersive nature of  tropospheric delays require a more careful
analysis of the data. We show how GPS data is processed to obtain the 
tropospheric slant delays using the GIPSY-OASIS II software
and define the concept of pseudo-wet delays, which will be the
observables in the tomographic software.
We then discuss the inverse problem in the 3D stochastic
tomography, using simulated refractivity fields to test the
system and the impact of noise. Finally, we use data from the Kilauea 
network in Hawaii and a local 4x4x41-voxel grid on
a region of 400 Km$^2$ and 15 Km in height to produce 4D refractivity
fields. Results are compared with ECMWF forecast.


\section{Introduction}
One goal of ongoing research on the application of tomographic techniques
to the atmosphere using GPS signals 
is to obtain 4D images of the refractivity in the troposphere
using a ground network. In this paper we present our first results on the
subject, showing that a careful combination of GPS data processing together
with tomographic Kalman-filtering can reproduce the state of the refractivity in the
neutral atmosphere.

The effect of the atmosphere on GPS signals is measured as an extra delay. The ionospheric
electron content produces a dispersive delay and can hence be corrected by
using two different frequencies and linearly combining them. The neutral atmosphere, however,
induces a delay independent of frequency. This delay ($\Delta L$) can be related to the refractivity 
by 
\bea
\label{delay_eq} \Delta L =\int_L 10^{-6} Ndl,\\
\label{refract_eq} N\approx 77.6\frac{P}{T}+3.73\cdot 10^5\frac{P_w}{T^2},
\eea
where $P$ is the total atmospheric pressure (mbar), $P_w$ is the water
vapor pressure (mbar), and  $T$ is the atmospheric temperature (K) (\cite{Smith}). The
extraction of this slant delay from GPS measurements requires accurate 
modeling. To take into account the dependence of the
tropospheric delay on the satellite elevation, mapping functions are used,
and a Zenith Total Delay (ZTD) is estimated at each station. To consider
azimuthal variability, a horizontal gradient is also estimated. The
time evolutions of these parameters is modeled as random walk stochastic 
processes. The total atmospheric delay can be partitioned into two components: 
the {\it {hydrostatic delay}}, due to the dry gases in the troposphere and the nondipole component of water vapor,
and the {\it {wet delay}}, due to the dipole component of water vapor. The
contribution of the hydrostatic component to the ZTD is larger than that of the
wet and it can be estimated from
surface pressure measurements.  We remove
the zenith hydrostatic delay (ZHD) from the total slant delays through an accurate 
estimation of the pressure
according to site measurements, and then form the {\it pseudo-wet delays} (PWD): 
we map the time series zenith wet delay (ZWD) and total horizontal gradients back
to each ray direction and add the residuals to account for any data mismodeling.
The refractivity associated is termed pseudo-wet refractivity ($\hat N$). The PWD's 
are the observables in the tomographic approach.
The reason for removing of the ZHD is two-fold. 
On the one hand, it
reduces the scale height of the problem (typically, the hydrostatic component has a scale
height of 10 Km while the wet component has a 2 Km scale height) and thus eases the numerical solution. 
In addition, the $\hat N$ field obtained 
is closely related to the water vapor distribution. 

In the paper we will first discuss the approach taken to estimate the 
PWD's using the JPL software GIPSY-OASIS II (GOA II) (see \cite{webb})
and the
network used in the study. Then we briefly discuss the inverse problem
and analyze the impact of the needed constraints through simulations. 
Finally, tomographic solutions with GPS data measured in the Kilauea network in Hawaii
during 1st February 1997 and their comparison with independent
measurement sources will be presented.
\section {Tropospheric Slant Delay Estimation}
The GPS signals $L_1$ and $L_2$ are first combined to remove the effect of
the ionosphere. The resulting observable, $L_C$, is modeled in GOA II using 
a Point Positioning Strategy in terms
of tropospheric effects (ZWD+ZHD+Gradients), geometrical factors and receiver 
clocks.
Satellite clock corrections and orbits are provided by JPL, as well as earth
orientation parameters. Tropospheric delays are mapped to the zenith using
Niell mapping functions (\cite{Niell}) and related to the gradients following the expression (\cite{Davis}):
\bea
D_L&=&m_h(e)D_{hz}+m_w(e)D_{wz}+\nonumber \\
&+&m_\Delta (e)\cot e [G_N\cos \phi +G_E \sin \phi],
\label{delay_yoaz}
\eea
The time dependence of the tropospheric component is handled using a kalman 
filter and a random walk stochastic process, with a drift rate of 
$0.25$ cm/$\sqrt{h}$ for the zenith delay and $0.03 $ cm/$\sqrt{h}$ (\cite{yoaz}) for the gradients. 
We then form the PWD from the solution as discussed above.
\subsection{Network and data considered}
\begin{figure}[h!]
\hspace{0.0cm} \epsffile{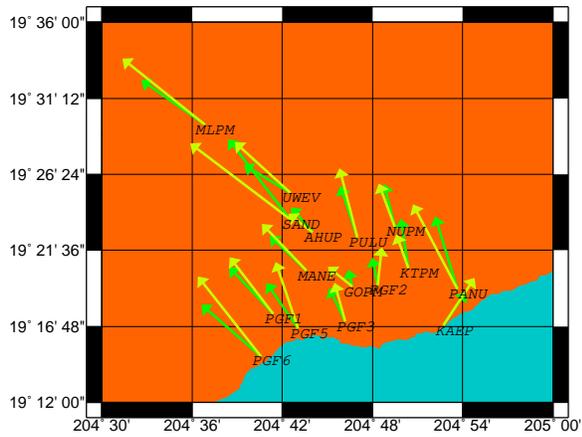}
\caption{ 24-h Reconstructed horizontal gradients (green) compared to GOA II solutions (yellow). Latitude in y-axis and Longitude in x-axis.}
\label{results1}
\end{figure}

GPS data was tracked with a sampling of 30 s in the Kilauea network in 
Hawaii. A map with the stations is shown in Figure \ref{results1}. 
It can be seen that they cover an area
of about 400 Km$^2$. The heights of the stations are distributed in a range
of about 2000 m, which  seems particularly well-suited for vertical
resolution in the tomographic solution.
\section {Tomographic approach and results}
\subsection{The inverse problem}
In tomography, one wants to obtain the solution 
fields (refractivity in the tropospheric case) from the integrated value 
along a ray path. The amount and distribution of  data, however,
 is often insufficient to determine a unique solution. Thus, some additional
information has to be added to the system. To this end, we rewrite the
pseudo-wet refractivity as 
$ \hat N(\vec r,t)=\sum_J a_J(t) \Psi_J(\vec r) + \epsilon(\vec r, t)$
where $\Psi_J(\vec r)$ is any 
set of basis functions and $\vec \epsilon(\vec r, t)$ is the quantization 
error. In our tomographic approach, three dimensional voxels
are used. Then, our observations are modeled by $y_i=T_i=
\int_{s.l.}\hat N(\vec r,t) dl=\sum_Ja_J(t)\int_{s.l.}\Psi_J(\vec r) d\vec l +\xi(\vec r,t)$.
 This is a set of linear equations of the form $Ax=y$. As mentioned, the
system may not have a solution and thus, we seek to
minimize the functional $\chi^2(x)=(y-Ax)^T\cdot (y-Ax)$. Even so, the
unknowns may not be fixed by the data, although the number of equations
is usually much greater than the number of unknowns. This is so because we are trying
to obtain a solution with more degrees of freedom (more voxels) than the
resolution carried by the data.
In previous work, (see \cite{Ruffini} and \cite{Rius}) we have discussed the
use of a correlation functional to confine the spatial spectrum of the
solution to the low portion of the frequency space. The same concept can
be expressed by adding new equations ({\it constraints}) that impose that the
density in a voxel be a weighted average of its neighbours.
Let us now take a closer look to the constraints. Horizontal smoothing
is needed to account for the non-uniform distribution of the
rays. Water vapor 
 is mostly concentrated in the first 2 Km in height.
Now, if we aim for a tomographic solution with a resolution of 350 m
(a trade-off between data resolution and computer load),  
some smoothing constraints have to be added in the vertical direction.
Finally, we impose a zero value constraint in the highest layer. 
The constraints are represented by the equation $B\cdot x =0$. To
smooth out the time variability, a Kalman filtering was used, modeling the 
troposphere as a random walk stochastic process with a drift rate of $\delta=0.14$  cm/(Km $\sqrt{h}$). The tomographic process is integrated in our home-made 
software package LOTTOS (LOcal Tropospheric TOmography Software).
\subsection {Results with simulated data}

\begin{figure}[h!]
\hspace{0.0cm} \epsffile{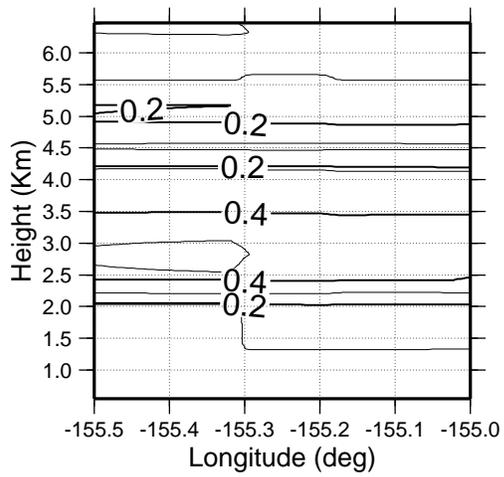}
\caption{ Simulation Results: Error (in contours, labeled at 0.2 cm/Km steps) in the reconstruction
as a function of longitude (in degrees East), 
for a constant latitude.}
\label{error}
\end{figure}

In order to test the software for the network's geometry and tune the different 
parameters, simulations play an essential
role. We have simulated a 3D pseudo-wet refractivity field following the 
expression (following \cite{Davis})
{\small
\bea \label{ref_eq}
\hat N(\vec \rho,z)=N_0^We^{-\frac{z}{h_{w}}}(1+\vec g_w\cdot \vec \rho)+N_{0r}^D e^{-\frac{z}{h_{d}}}(1+\vec g_d\cdot \vec \rho)
\eea
}where $h_{w}=2$ Km, $h_d=10$ Km, $\vec g_d$ and $\vec g_w$ are the horizontal gradients
up to a multiplicative constant in the
hydrostatic and wet component, respectively and $\vec \rho$ is the horizontal coordinate.
We have set $N_{0}^W=150$ mm/Km and $N_{0r}^D=2$ mm/Km, the latter to account for any residual 
hydrostatic component, $\vec g_d=0$ and two different
non-zero $\vec g_w$ values, applied depending on the latitude of the station. 
The geometry of the rays for the 1st February 1997 in the Kilauea network has been
used to generate the simulated rays. The PWD for each ray has been formed according to
 $L_{sim}=\sum_{vox}\hat N_i\cdot l_i$,
where $\hat N_i$ is the pseudo-wet refractivity associated to voxel $i$ and $l_i$ is the length of the ray across
this voxel. The weight of the constraints was tuned to give the best fit with
simulations, successively adding the constraints in the horizontal plane,
then the vertical smoothing constraints, and finally adding a zero value constraint for
the highest layer. In Figure \ref{error} we show the error in reconstructing 
the simulated field.

Vertical reconstruction has also been verified adding a perturbation at different
heights to the simulated fields of Equation \ref{ref_eq};
good vertical resolution is achieved  thanks to
the distribution in height of the stations. We also note (see Figure \ref{error})
 that vertical reconstruction is gradually lost as the perturbation is
placed at higher altitudes (above 2 Km, resolution starts degrading). However, 
when the boundary constraint kicks in, the solution agrees again with the simulated
field, slowly decaying to zero.

\begin{figure}[h!]
\hspace{0.0cm} \epsffile{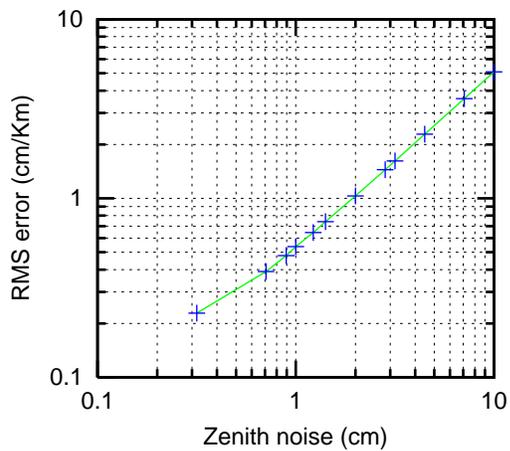}
\caption{Impact of white noise in the tomographic solution.}
\label{simulation2} 
\end{figure}

The impact of noise in the rays has also been considered. Following the data
processing described above, we have seen that post-fit residuals are dependent on
elevation (roughly as  1/sin(e), where $e$ is the elevation); therefore, 
we can assume a noise value for
the ZWD, then map it to the slant directions, and add it to the total simulated
delay. We show in Figure \ref{simulation2} the results of the noise analysis.
As it can be seen, a ZWD noise of 0.5 cm represents a noise in the solution
of less than 3 mm/Km rms (about 2\% of the maximum value).

\subsection {Results from the Kilauea network}
\begin{figure}[h!]
\hspace{0.0cm} \epsffile{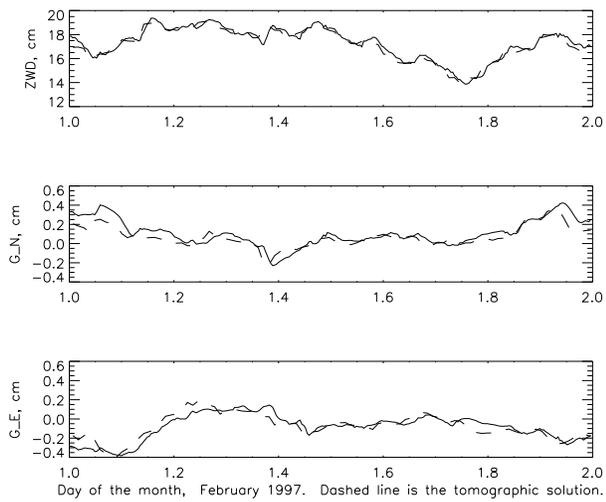}
\caption{ Time series of the ZWD and Gradients as computed
using GOA II (solid) and the tomographic fields (dashed) for PGF3.}
\label{results2}
\end{figure}

\begin{figure}[h!]
\hspace{0.0cm} \epsffile{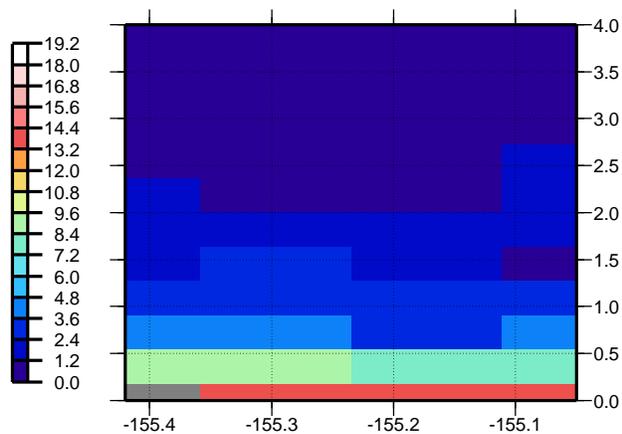}
\caption{Image of the refractivity field for a latitude of 19.28$^o$ N, February 1st, in the Hawaii network, having Longitude in x-axis (degrees East) and height
(Km) in y-axis. Color scale is in cm/Km.}
\label{results3} 
\end{figure}

\begin{figure}[h!]
\hspace{0.0cm} \epsffile{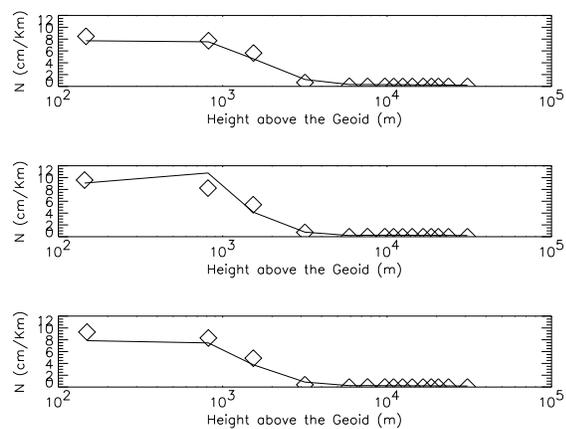}
\caption{ Comparison of tomographic solution (solid line) with ECMWF (diamonds)
for hours 06h (top), 12h (middle) and 18h (bottom). }
\label{results4}
\end{figure}

Pseudo-wet delays from the Kilauea network in Hawaii have been input to the system.
We have processed data from 18 stations for February 1st, 1997.
We have used LOTTOS considering 4x4 voxels of $6'$ in latitude and $7'$ 
in longitude, 41 layers of $350$ m height 
and 30-minutes batches for the Kalman filter. The tomographic fields have then been
used to generate the reconstructed PWD which, in turn, have been processed to calculate
the horizontal gradients and ZWD for each station. These have been compared
to those obtained with GOA II.
In Figure \ref{results1} we show the magnitude and direction of the 24-h mean value of the gradient as calculated
using GOA II (yellow) and using the tomographic solution (green).
In Figure \ref{results2} we show the time series of these magnitudes for
a particular station (PGF3) and  in Table 1 we show the correlation values
over time for each station. This shows that the reconstructed 4D field  is 
consistent with the data. Vertical distribution is shown in Figure \ref{results3} 
for a given latitude and time. We have computed the values of $\hat N(\vec r,t)$
profiles with data from the European Center for Medium-Range Weather
Forecasts (ECMWF) analysis and compared with our tomographic profiles,
averaging them to meet the ECMWF maps resolution of 0.5 degrees.
Results are presented for three different hours (06h, 12h and 18h) in 
Figure \ref{results4} showing good agreement.

\section {Conclusions}

We have successfully obtained tomographic 4D images of the pseudo-wet
refractivity for a local dense network. The concept of slant pseudo-wet delays
has been introduced. Simulated fields have been used to validate and
tune our LOTTOS software package. Finally, data from the Kilauea
 network have been 
processed and the reconstructed 4D fields of the refractivity have been
compared with on-site estimated horizontal gradients and ECMWF vertical
profiles, yielding good correlation in both cases. Although more work is
needed in this area, our results provide the proof-of-concept
of the tomographic inversion of tropospheric GPS data.
\section*{Acknowledgements}
\small We thank the USGS Hawaii Volcano Observatory, 
Stanford University and the University of Hawaii, and JPL for providing the data.
 We thank Yoaz Bar--Sever for
useful conversations and Beatriz Navascues (Instituto Nacional de Meteorologia)
 for providing the ECMWF data. This work was supported by Spanish Climate CICYT
 grant CLI95-1781, EC grant WAVEFRONT PL-952007 and the Comissionat per a 
Universitats i Recerca de la Generalitat de Catalunya.
\newpage
 
   %
   %
   %

%
%

   %
   %
   %

{ \renewcommand{\baselinestretch}{1.0} \small  
\begin{tabular}{|c|c|c|c|}
\hline
{\bf Station} & {\bf ZWD} &  {\bf North } & {\bf East} \\
\hline 
AHUP & 0.98 & 0.80 & 0.83 \\
GOPM & 0.96 & 0.79 & 0.64 \\
KAEP & 0.98 & 0.78 & 0.71 \\
KTPM & 0.97 & 0.57 & 0.72 \\
MANE & 0.98 & 0.90 & 0.70 \\
MLPM & 0.99 & 0.55 & 0.88 \\
NUPM & 0.97 & 0.62 & 0.53 \\
PANU & 0.99 & 0.53 & 0.43 \\
PGF1 & 0.98 & 0.83 & 0.83 \\
PGF2 & 0.97 & 0.74 & 0.78 \\
PGF3 & 0.97 & 0.86 & 0.86 \\
PGF5 & 0.97 & 0.85 & 0.84 \\
PGF6 & 0.97 & 0.88 & 0.89 \\
PULU & 0.98 & 0.56 & 0.41 \\
SAND & 0.99 & 0.61 & 0.82 \\
UWEV & 0.99 & 0.69 & 0.82 \\
\hline 
\end{tabular} 
}
\vspace{1 cm}

\noindent {\bf Table 1}: 24-h correlation factors for
the ZWD \\
and  the Total Horizontal Gradients in both \\
components, North and East, for each station.

\end{document}